# Influence of limestone filler and of the size of the aggregates on DEF

M. Al Shamaa[1], S. Lavaud[1], L Divet[1], JB Colliat[2], G Nahas[3], JM Torrenti[1]

1) Université Paris-Est, IFSTTAR, 14-20 boulevard Newton, cité Descartes, Champs sur Marne, F-77455, Marne la Vallée, France
2) Laboratoire de Mécanique de Lille, Université Lille 1 Sciences et Technologies, CNRS, Ecole Centrale de Lille, Arts et Métiers ParisTech, Boulevard Paul Langevin, Cité Scientifique, 59655 Villeneuve d'Ascq Cedex, France
3) Institut de Radioprotection et de Sûreté Nucléaire (IRSN), 31 avenue de la Division Leclerc, 92260
Fontenay-aux-Roses, France

Abstract

This experimental study aims to determine the effect of limestone filler on concrete expansion due to delayed ettringite formation (DEF). Different mortars made with different sizes and percentages of limestone filler and Portland cement CEM I 52.5N are conserved in water. The expansion of the specimens is measured. Results show that DEF is not inhibited by limestone filler. The kinetics and the amplitude of the swelling depend on the size of the limestone filler. The volume fraction of aggregates changes only the kinetics: the relation between swelling and water uptake depends only on the size of the aggregates

1 introduction

Delayed ettringite formation (DEF) in concrete structures is a pathology that can develop in concrete where temperature at early age is high. It is mainly found in structures that are in contact with a moist environment for several years [7] but also in the case of endogenous conditions [2]. The binder's composition is a major parameter of this phenomenon: DEF could be prevented when a part of the cement is replaced by pozzolanic or latent hydraulic additions,[9, 11]. In case of using limestone filler, the results are more contrasted: Silva et al. [12, 13] have shown that DEF was not inhibited by this type of additions (in fact they found a larger swelling when at least 10% of the cement was replaced by limestone filler), while Kurdowski and Duszak [15] have found that limestone presents the same efficiency as fly ash. Silva states that this belongs to a chemical effect: when slag is used, the consumption of portlandite by the pozzolanic reaction or the lower created quantity of portlandite affect the alkalinity of the interstitial water, thus inhibiting the formation of ettringite. This effect is not present in case of using limestone filler. In the work of Kurdowski, the temperature cycle was representative to that used in prefabrication (6 hours at 90°C), while in Silva's work it was representative to that occurring in a massive structure (temperatures above 70ºC during 3 days). In addition, the duration of the plateau at high temperatures (larger than 65°C) is a major parameter for the magnitude of swelling due to DEF [6].

The aggregate size seems to be an important parameter as well: Fu et al. [16] have shown that the smaller the aggregate size, the larger the concrete swelling. Grattan-Bellew et al. [17], with siliceous aggregates (quartz), observed that the expansion is inversely proportional to the aggregate size and that the rate of expansion increased rapidly as the mean particle size decreased. In addition, Heinz et al. [18] have shown that the latency time decreases when the size of the aggregates increases. The explanation is mechanical: in case of aggregates with large dilation coefficient and when concrete is subjected to high temperatures at early age, cracks are induced thus leading to foster the swelling. These cracks are larger when the size of aggregates increases. But this seems to be only a delayed effect because in case of pure cement paste, DEF was also observed but with a very long latency time [14]. It should be noted that in the work of Kurdowski, the size of the limestone filler was not indicated while in Silva's work the mean diameter of the limestone aggregates is 7 μm.

In this work, two questions were discussed by using limestone filler with different size of aggregates and a temperature cycle representative to a massive structure: we discuss the effect of the replacement of cement by limestone filler and the influence of the size of the aggregates on the magnitude of swelling of mortar samples.

2 experiments

2.1 materials

The compositions of the mortars used in this study are given in Table 1. The mortars are made with different sizes and percentages of limestone filler and Portland cement CEM I 52.5N. The chemical and mineralogical compositions of the cement are given in Tables 2 and 3. This cement was selected because of its large sulfate and alkali contents, which are major factors for DEF [19]. It was already used in a previous study [2].

Four sizes with mean diameter of 20 µm, 45 µm, 510 µm and 2800 µm, and three percentages of 30, 40 and 50% were selected for this study. The nomenclature used for the mortar samples was *dxpy* where *x* and *y* correspond to the mean diameter and the percentage of filler respectively. The chemical composition of the aggregates is the same (table 4) and the only difference between them is the mean diameter (Figure 1).

The quantity of water of each mortar was adjusted in order to obtain the same workability (slump between 7 and 10 cm by using the mini cone test). It should be noted that the water-cement ratio for the filler having a size of 20 and 45 µm is considered as an approximation due to fact that the water absorption measurement of these materials is not accurate.

| Reference | Cement (g) | Limestone filler (g) | Total water (g) | Absorbed water (g) | W/C |
|---|---|---|---|---|---|
| d20p30 | 19319 | 16200 | 12426 | 4698 | 0,40 |
| d20p40 | 16114 | 21600 | 13032 | 6264 | 0,42 |
| d20p50 | 12908 | 27000 | 13639 | 7830 | 0,45 |
| d45p30 | 20750 | 16200 | 10827 | 3564 | 0,35 |
| d45p40 | 17786 | 21600 | 10977 | 4752 | 0,35 |
| d45p50 | 13799 | 27000 | 11460 | 5940 | 0,40 |
| d510p30 | 20218 | 16200 | 7582 | 65 | 0,37 |
| d510p40 | 16726 | 21600 | 6691 | 86 | 0,39 |
| d510p50 | 10900 | 27000 | 6540 | 108 | 0,59 |
| d2800p30 | 20218 | 16200 | 7582 | 65 | 0,37 |
| d2800p40 | 16726 | 21600 | 6691 | 86 | 0,39 |
| d2800p50 | 13030 | 27000 | 5863 | 108 | 0,44 |

**Table 1. Mortars compositions** (dosage for 20 L).

| CEM I 52.5 | Wt.% |
|---|---|
| $SiO_2$ | 19.19 |
| $Al_2O_3$ | 5.03 |
| $Fe_2O_3$ | 2.06 |
| $TiO_2$ | 0.31 |
| MnO | 0.04 |
| CaO | 62.68 |
| MgO | 0.92 |
| $SO_3$ | 3.39 |
| $K_2O$ | 1.03 |
| $Na_2O$ | 0.12 |
| $P_2O_5$ | 0.25 |
| $Na_2O_{eq}$ | 0.77 |
| $S^{2-}$ | 0.03 |
| $Cl^{2-}$ | 0.02 |

**Table 2.** Chemical composition of the Portland cement

| CEM I 52.5 | Wt.% |
|---|---|
| $C_3S$ | 67.4 |
| $C_2S$ | 11.3 |
| $C_3A$ | 10.7 |
| $C_4AF$ | 7.3 |

**Table 3.** Mineralogical composition of the Portland cement

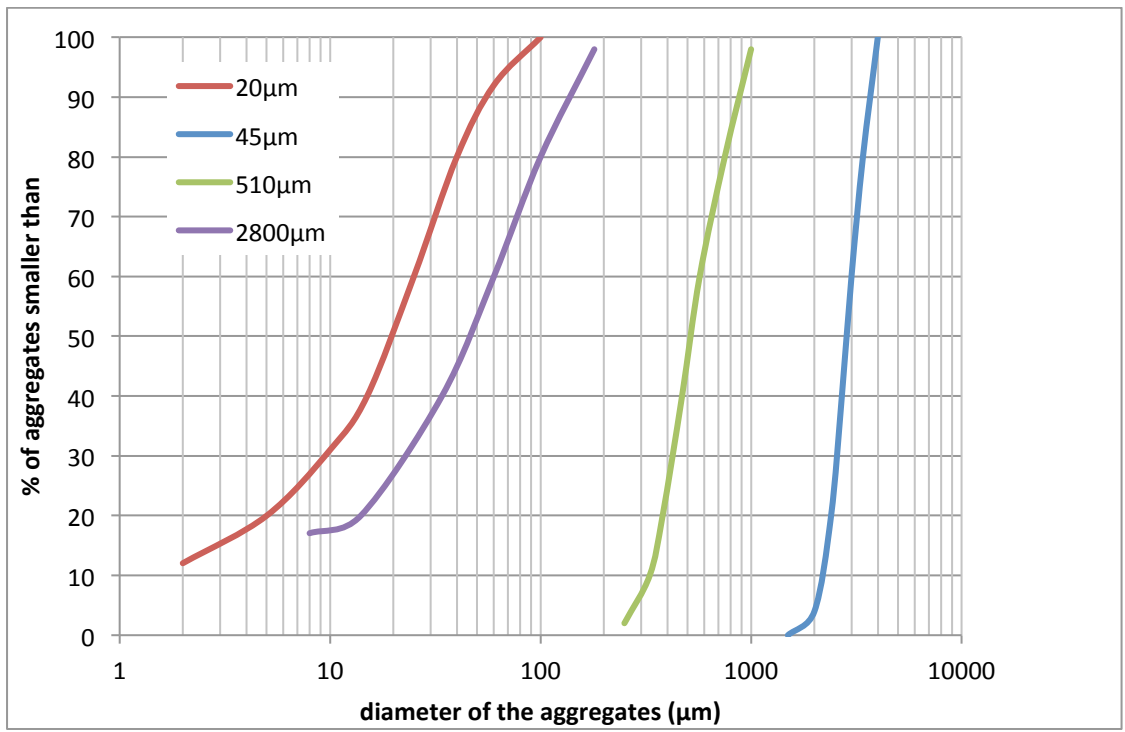

**Figure1.** Granular distribution of the aggregates.

2.2 specimens curing and conservation

Cylindrical samples with diameter of 11 cm and height of 22 cm were casted for each mortar. After casting, the samples were tightly covered to prevent evaporation of water and then cured in their moulds according to the heat treatment presented in Figure 2. The specimens were placed in a drying oven equipped with a temperature/humidity controller. The treatment is divided into four steps. Firstly, the samples were stored at 20°C for 2 hours. Then, the temperature was increased at a rate of 2.5°C/h in order to reach 80°C. The temperature was maintained at 80°C for 3 days. Finally, the temperature was decreased at a rate of -1°C/h in order to reach 20°C. So the total duration of the heat treatment is 7 days. This treatment is representative to the temperature cycle in a massive concrete structure due to the heat generated by the cement's hydration (see [1] as an example of the case of a nuclear power plant raft foundation). After the heat treatment, the specimens were demoulded and then stored at 20°C in a box filled with tap water (there was no immersion of different mortars in the same box).

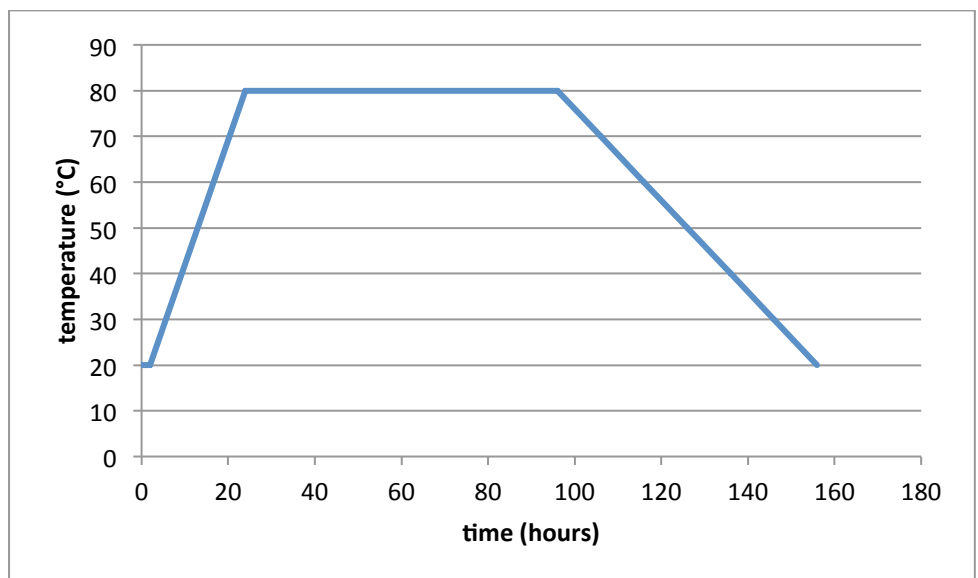

Figure2. heat treatment applied to the mortar samples

### 2.3 measurements

Mass uptake and swelling evolution were measured for each sample. Water uptake was monitored by weighting the specimens while the axial expansion was monitored with a digital extensometer by measuring the length change using steel studs glued on three locations placed at 120 degrees from each other around the diameter. These studs are aligned per pair along the height of the specimen.

The initial distance between two studs is 10 cm, and the variation of this distance over time assesses the expansion of the specimen.

These measurements were completed by Scanning Electron Microscope (SEM) observations of polished sections. Total porosity measurement and a quantitative analysis of the crack pattern were also conducted for the samples having the largest swelling. Water porosity measurements were carried out according to the norm NF P 18-459. The cracks analysis was conducted by impregnation of samples with fluorescent resin which increases the contrast between cracks and plain mortar. Slices (11cm of diameter and 2cm of thickness) were examined at 28 days and at 280 days by using a picture analysis program developed at Ifsttar. Figure 3 presents an example of this treatment. In order to minimize the effect of variability 100 images (size of each image 3mmx4mm) were treated for each slice. This analysis was performed at 28 and 280 days for the mortars d2800p30, d2800p40 and d2800p50. The quantification of the microcracking is made firstly by the means of the specific length of the cracks then using the secant method: the image is covered by a grid of parallel lines (so called directed secants) and the intersections with the cracks are counted for several directions of the secant lines [20]. This measurement allows the quantification of orientation of the cracks.

a.
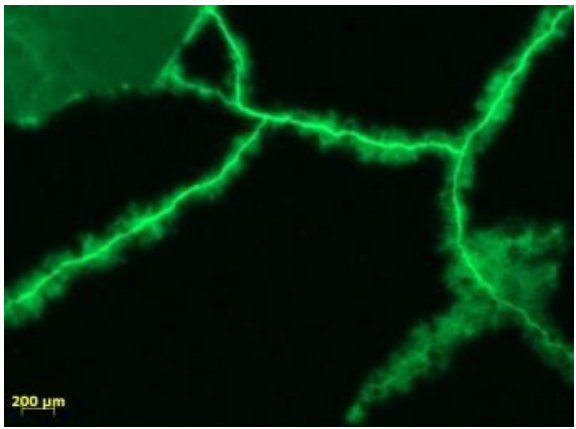
c.

b.
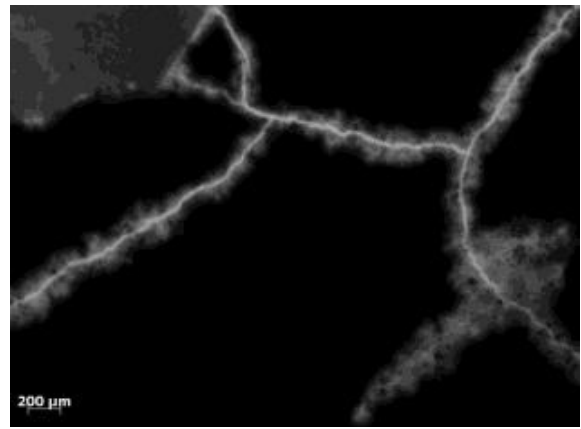
d.

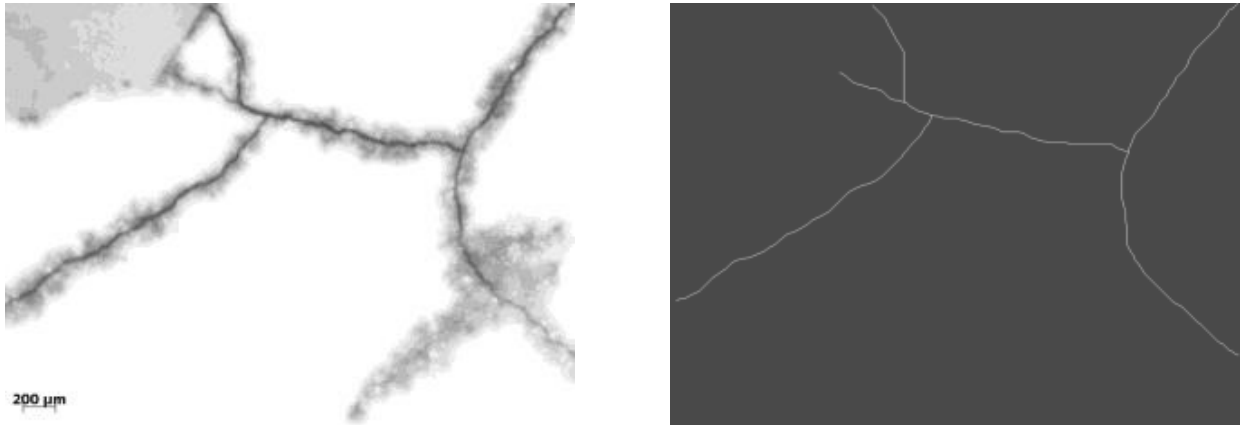

Figure 3: example of image analysis (size of the image = 3mm x 4mm): a. direct view of the slice under the fluorescence microscope b. grayscale image c. reversed grayscale image d. segmented image after application of a level set method.

3 Analysis of the experimental results

3.1 Mass uptake and swelling

Figure 4 shows the expansion evolutions for all the mortars (except for specimen d20p50 for which the measurements were stopped after one year because of a removal of the studs; at that time no swelling was observed). The results show that all the samples present swelling and that the larger the aggregate size the larger the swelling. Expansion was also increased with the limestone filler percentage. In addition, expansions versus weight variations showed a unique hydrous behaviour for each aggregate size (Figure 5 and 6; the curves are too much perturbed for the smallest aggregates to be presented due to the low swelling).

This indicates that the mortars behaviour is only affected by the size of the limestone filler aggregates, the percentage of limestone playing a role on the swelling kinetic only.

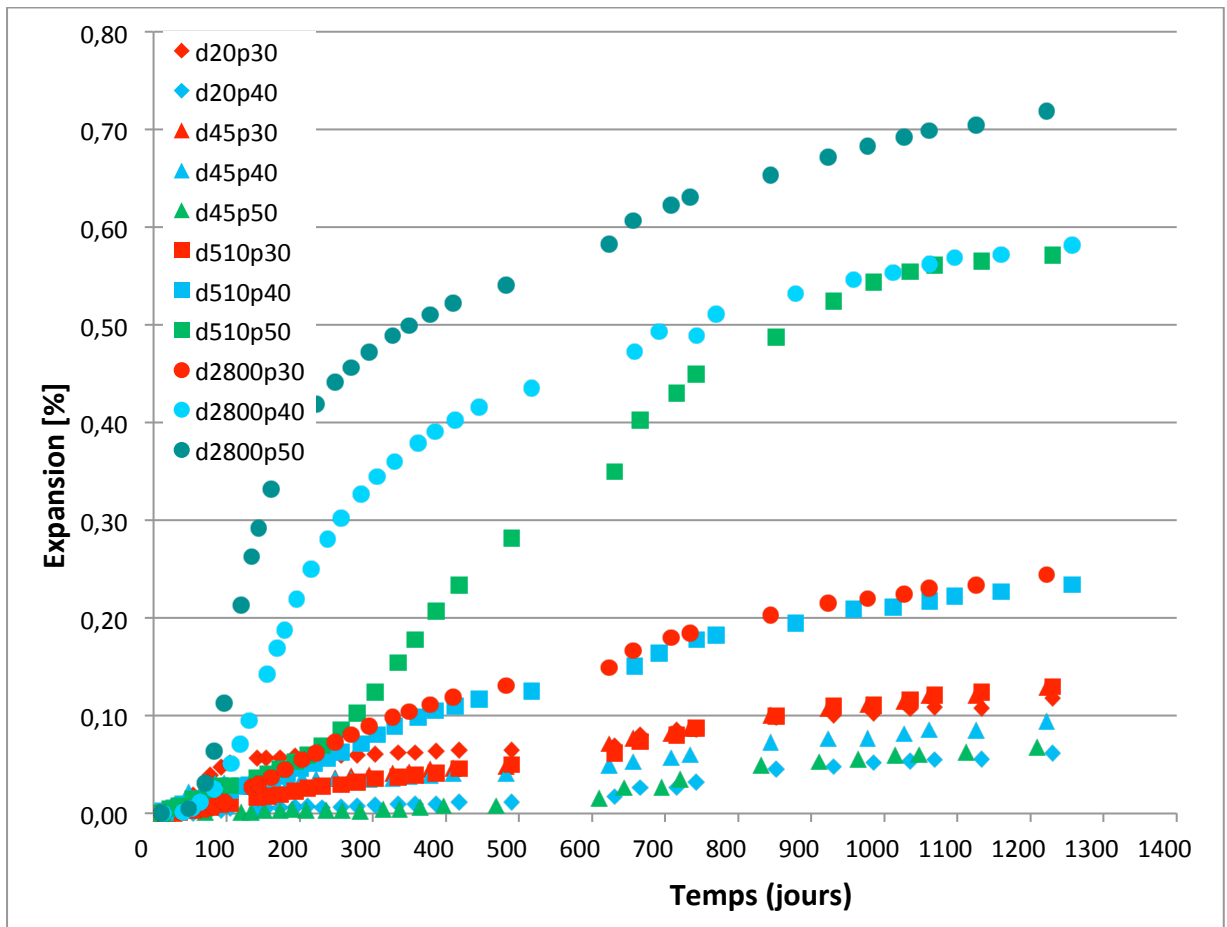

Figure 4. Evolution of swelling of all specimens (except specimen d20p50)

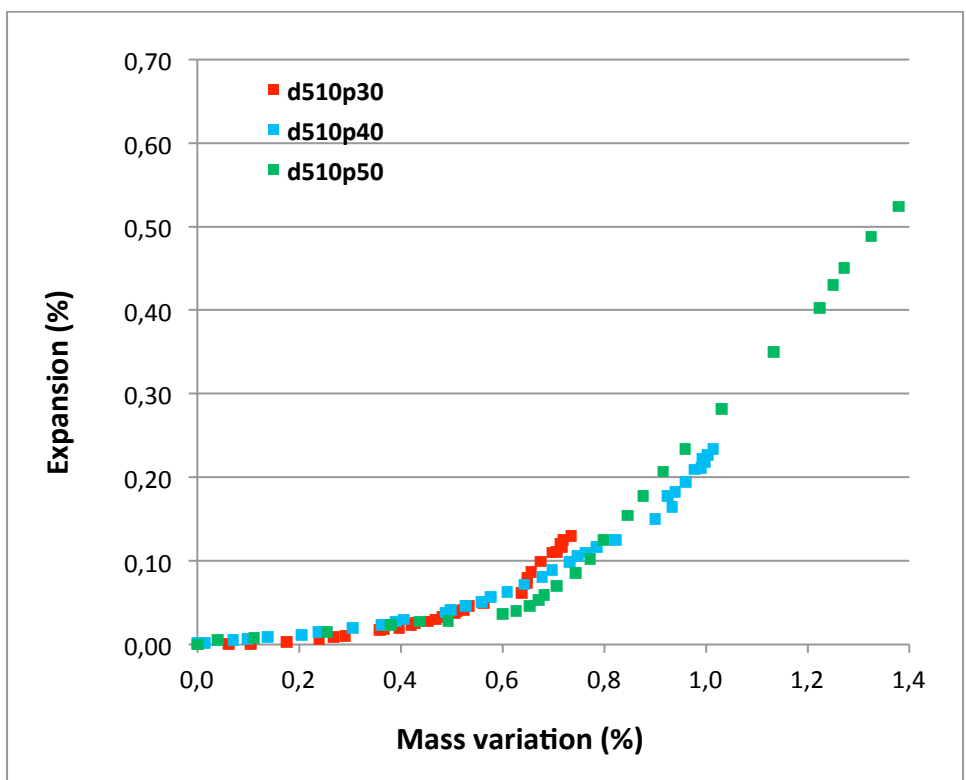

Figure 5. Evolution of the swelling with the mass uptake – case of 510μm aggregates.

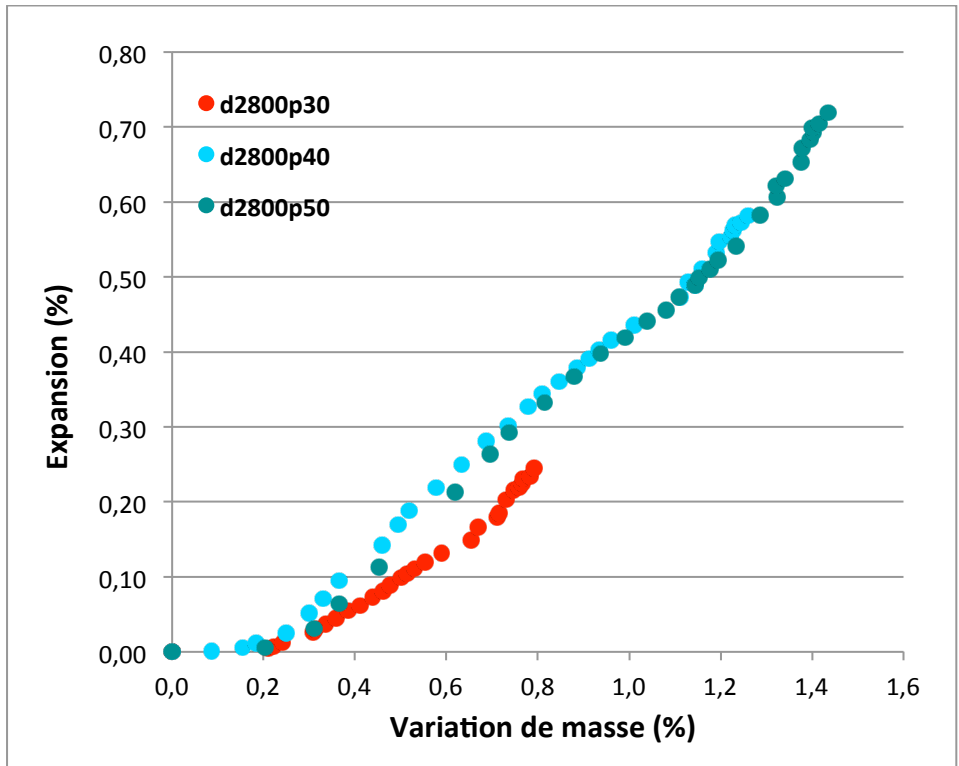

Figure 6. Evolution of the swelling with the mass uptake – case of 2800μm aggregates.

3.2 Porosity

Table 4 presents the total water porosity of the samples with the largest aggregates. It can be seen from figure 4 that samples with the same porosity at 28 days (like for instance d510p40 and d2800p40) have rather different behaviors. Porosity is certainly a parameter for the development of DEF but less important than the size of the aggregates. A small increase of the porosity could also be noticed between 28 and 280 days, indicating that the creation of new cracks after swelling is larger than the filling of the existing pores by ettringite.

|  | d510p30 | d510p40 | d510p50 | d2800p30 | d2800p40 | d2800p50 |
|---|---|---|---|---|---|---|
| 28 days | 25,5% | 22,5% | 24,0% | 25,5% | 22,3% | 20,2% |
| 280 days | 26,1% | 23,7% | 25,0% | 25,5% | 24,0% | 21,0% |

Table4. Porosities of d510py and d2800py samples.

3.2 SEM observations and quantitative analysis of the crack pattern

Despite the fact that SEM observations were only qualitative, ettringite was clearly visible in specimen d2800p50 and d510p50 (figure 7) at the interface between the aggregates and the cement paste and within cracks in the cement matrix. In specimen d2800p30 and d510p30, ettringite was not visible at the interface and could only be seen in the form of nodules (figure 8).

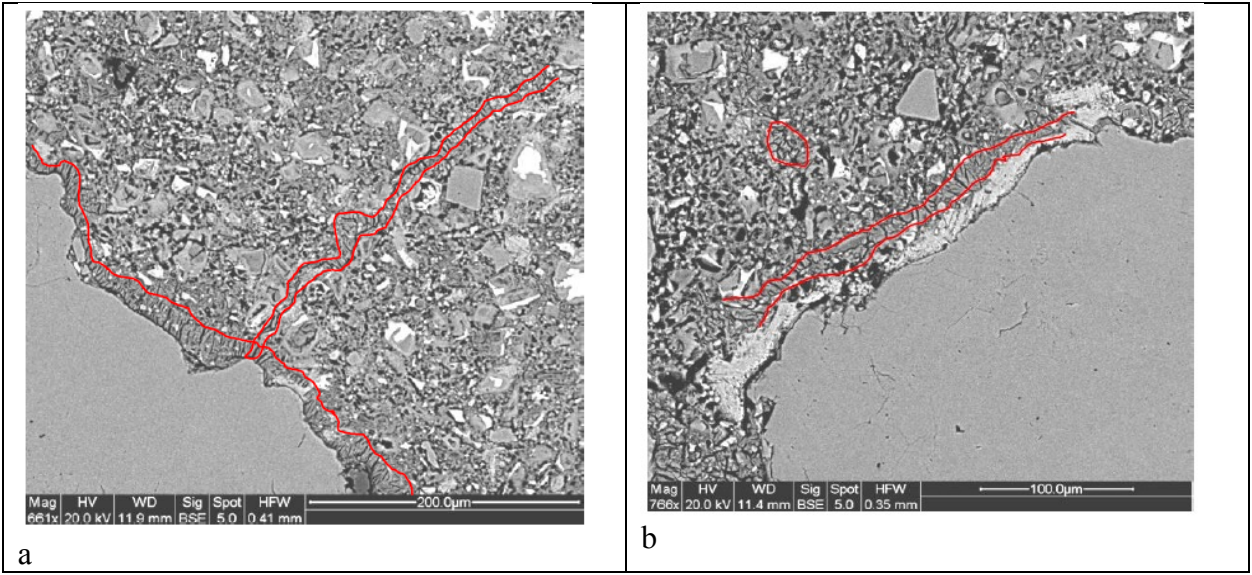

Figure 7: a) d2800p50 b) d510p50; ettringite is located in the area delimited by the red lines

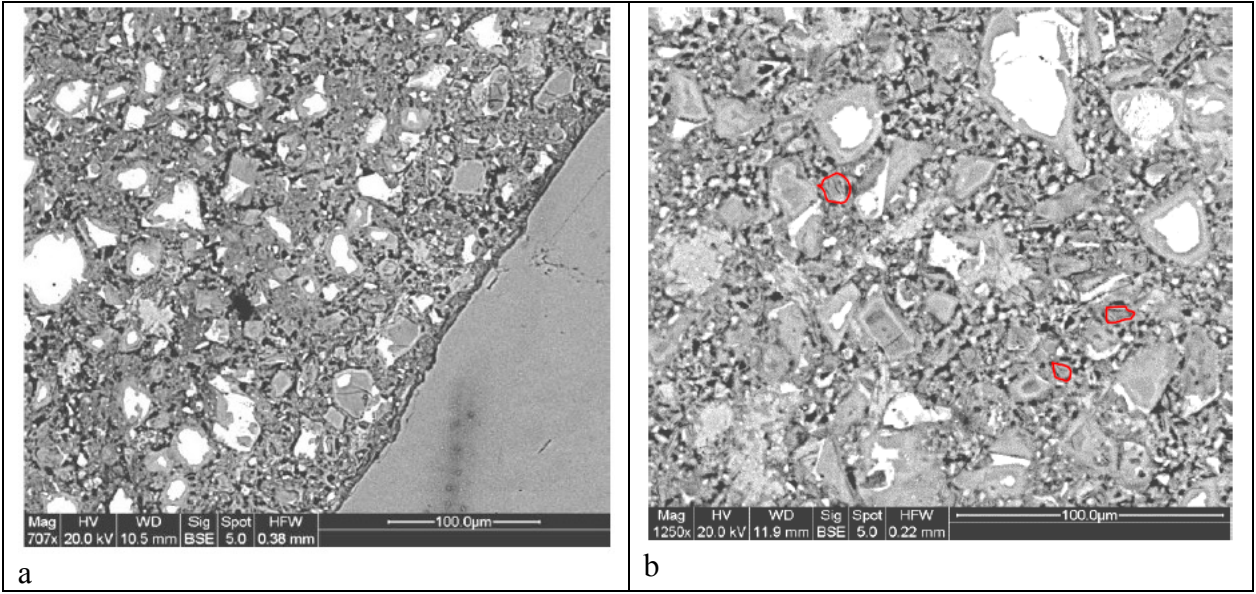

Figure 8: d2800p30 a) ettringite is not visible at the interface b) small nodules of ettringite visible within the matrix

Figure 9 presents the evolution of the total length of the cracks. The increase of the microcracking due to DEF is clearly visible on all the assessed samples. The influence of the aggregate size is also noticeable. Concerning the orientation of the cracks, the obtained results show a rather isotropic distribution of the directions of the cracks.

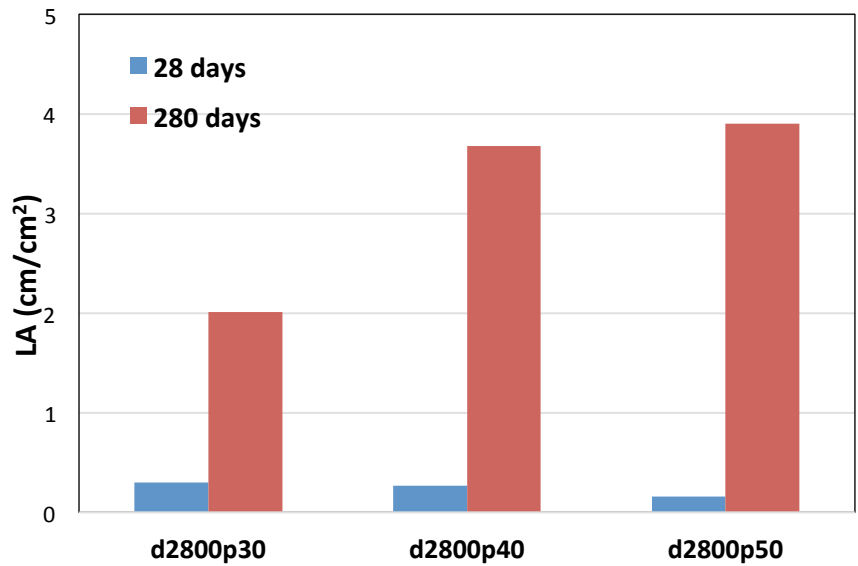

Figure 9: evolution of the total length of the cracks for 3 different mixes (in blue=at 28 days, just after heat treatment, in red=at 280 days).

3.3 Analysis and discussion of the results

All the mortars that were studied have presented swelling. Despite a possible reaction of dissolved carbonate ions from the limestone filler with the aluminate phases of Portland cement which leads to the formation of carboaluminates as opposed to sulfoaluminates, and which stabilizes the ettringite that is produced at early ages [4, 8], limestone filler clearly does not inhibit delayed ettringite formation.

The size of the limestone aggregates is a major parameter for the swelling of our specimens. When the size increases, the swelling increases and the latency time decreases. This could be explained on a chemical point of view and on a mechanical one. As mentioned previously, the consequence of a possible reaction between carbonate ions and aluminates will be the formation of carboaluminates and a lower quantity of aluminates available for the formation of ettringite. This effect is more important for the smaller particles because the surface area of limestone filler is larger [4]. The quality of interfacial transition zone is also better [4] and this could influence the transfer properties of the mortar.

On a mechanical point of view, cracking could be induced during the thermal treatment. This is due to different elastic characteristics and differential dilation coefficients between the cement paste and the limestone filler. When the thermal treatment is applied, tensile stresses are generated at the interface between the cement paste and the aggregates which lead to development of cracking. This initial cracking is visible (figure 9) and exists since the beginning. If we consider a single aggregate in a cement paste, the stress at the interface does not depend on the size of the aggregates [5]. But the perturbed zone depends on the size. This explains why the fracture process zone depends on the size of the aggregates [3] and why the crack opening is larger when the size of the aggregates increases. It finally facilitates the possibility of ettringite precipitation at the interface (figure 8).

The evolution of swelling with the mass uptake (figures 5 and 6) shows that, for a given size of the limestone filler, same behaviors are noticed. However, for a smaller percentage of filler, the mass uptake is lower and thereby the swelling. A possible explanation for this phenomenon is the fact that the connection between the interfacial transition zones (ITZ) depends on the percentage of aggregates. Due to interaction between the aggregates and the cement paste, ITZ has a larger porosity and thus allows a faster diffusion of ions. In the case of DEF, diffusion of alkali ions is needed for ettringite precipitation. Garboczi [21] has shown that with 30% of aggregates the percentage of connected ITZ is low. He has also shown that when the size of the aggregates decreases a larger percentage of aggregates is needed in order to connect the ITZ. This could also explain the effect of the size on swelling.

4 Conclusions

In this study mortars made with limestone filler as aggregates were prepared. Different mean sizes and percentage of aggregates were used. After a heat treatment corresponding to the temperature cycle of a massive structure, the samples were immersed and their evolutions were followed.

All the mortars that were studied have presented swelling: limestone filler clearly does not inhibit delayed ettringite formation. The kinetics and the amplitude of the swelling depend on the size of the limestone filler. The volume fraction of aggregates changes only the kinetics: the relation between swelling and water uptake depends only on the size of the aggregates.

The formation of carboaluminates could explain this behavior and its relation to the surface area of the limestone filler. The thermo-mechanical behavior could also explain the occurrence of cracking at the interface between aggregates and cement paste which later favorites the formation of ettringite within the cracks.

The results presented here could be used with a mesoscopic approach in order to model the delayed ettringite formation, see for instance [10].